*Article*

# Perceptual evaluation of Acoustic Level of Detail in Virtual Acoustic Environments

**Stefan Fichna [1]\*, Steven van de Par[2], Bernhard U. Seeber[3], Stephan D. Ewert [1]**

[1] Medizinische Physik and Cluster of Excellence Carl von Ossietzky University Oldenburg, 26129 Oldenburg, Germany; stefan.fichna@uol.de

[2] Akustik and Cluster of Excellence Hearing4all, Carl von Ossietzky University Oldenburg, 26129 Oldenburg, Germany; Universität Oldenburg; stephen.van.de.par.@uol.de

[3] Audio Information Processing, Technische Universität München, 80333 München; seeber@tum.de

\* Correspondence: stefan.fichna@uol.de

**Abstract**

Virtual acoustic environments enable the creation and simulation of realistic and ecologically valid daily-life situations vital for hearing research and audiology. Reverberant indoor environments are particularly important. For real-time applications, room acoustics simulation requires simplifications, however, the necessary acoustic level of detail (ALOD) remains unclear in order to capture all perceptually relevant effects. This study examines the impact of varying ALOD in simulations of three real environments: a living room with a coupled kitchen, a pub, and an underground station. ALOD was varied by generating different numbers of image sources for early reflections, or by excluding geometrical room details specific for each environment. Simulations were perceptually evaluated using headphones in comparison to binaural room impulse responses measured with a dummy head in the corresponding real environments, or by using loudspeakers. The study assessed the perceived overall difference for a pulse stimulus, a played electric bass and a speech token. Additionally, plausibility, speech intelligibility, and externalization were evaluated. Results indicate that a strong reduction in ALOD is feasible while maintaining similar plausibility, speech intelligibility, and externalization as with dummy head recordings. The number and accuracy of early reflections appear less relevant, provided diffuse late reverberation is appropriately represented.

## 1. Introduction

Room acoustics simulation and virtual acoustics are widely applied, ranging from generating highly controllable complex acoustic environments for hearing research and audiology [1-4], to entertainment and video game sound design [5,6], augmented reality (AR) or virtual reality (VR) [Picinali et al., 2023], as well as architectural planning [7]. Several room acoustics simulation approaches exist [8-14], each involving simplifications of the underlying acoustical processes. Many algorithms [10,11] simulate early reflections using the image source model (ISM) [15], combined with an alternative method to simulate late reverberation, such as ray tracing [16], feedback delay networks [17] extended to spatial rendering [11, 13, 18], scattering delay networks [19] and acoustic radiance transfer [14, 20]. State-of-the-art methods achieve a high degree of perceptual plausibility [21,22], defined as "a simulation in agreement with the listener's expectation towards a corresponding real event" as defined by Lindau and Weinzierl [23].

With regard to interactive real-time applications it is of particular interest to reduce the computational demands of room acoustics simulation while maintaining perceptual plausibility and achieving a certain agreement of spatial audio quality and speech intelligibility (SI) between simulation and the respective real-life environment. For this, the acoustic level of detail (ALOD) of the simulated environment might be reduced if the perceptual consequences remain small. The term ALOD is inspired by level of detail (LOD) as commonly used in computer graphics, to denote typically distance-dependent reductions in visual geometric model complexity (polygon count) while maintaining the same perceptual impression. Prior studies have examined the influence of ALOD on acoustic parameters such as reverberation time and the speech transmission index [24]. Abd Jalil et al. [25] demonstrated that the number of surfaces in acoustic models can be reduced by up to 80% without compromising accuracy. Moreover, it has been shown that even when simplifying more complex room shapes to a proxy rectangular ("shoebox") geometry [13,11,18], it is possible to maintain perceptual plausibility and achieve close alignment with measured reference conditions [21,26,22]. This level of simplification not only facilitates low-latency, interactive virtual acoustic environments for hearing research but also allows for the possibility of running realistic spatial rendering on mobile hardware, which is advantageous for fitting hearables and hearing support systems [27,28]. To ensure broad applicability of such simulations, it is important to evaluate perceptual outcomes using both headphone-based and loudspeaker-based reproduction methods. While headphone reproduction allows precise control over binaural cues, loudspeaker setups may be more suitable in scenarios where head tracking is unavailable or when simulating open-fit hearing devices. [4] have shown that shoebox simulations can yield good agreement with real-room references in terms of SI under both rendering approaches. However, it remains unclear whether certain simplifications in the acoustic model, such as omitting nearby reflective surfaces or coupled volumes, impact perception differently depending on the reproduction system.

Fichna et al. [29] investigated the perception of simulated and real acoustic scenes with varying ALOD using headphones. Their preliminary study examined SI in conditions with a masking signal and evaluated the overall perceived spatial audio quality using different ALOD in the acoustic simulation in comparison with recordings for a pink-colored pulse as a target signal. Fichna et al. [30] extended this research by systematically evaluating plausibility, the overall perceived difference between the renderings, and the perceived externalization with headphones and with loudspeakers. They used speech as well as a pink-colored pulse as target sounds. Their preliminary data showed that simulations with a maximum ALOD were as plausible as those generated using measured binaural room impulse responses (BRIRs), even with pink-colored pulses, which generally provides the highest sensitivity for perceptual differences in room acoustics simulation [see, e.g., 18]. Despite high plausibility ratings observed in Fichna et al. [30], a considerable perceived overall difference between simulated and measured conditions remained. However, it remains unclear to which perceptual qualities these differences are related and conclusions are limited by the relatively small number of subjects in these studies.

Recently, [31] investigated implications of simplifying geometrical acoustics by reducing a highly resolved surface model of a real home environment down to a shoebox model. They used the CATT Acoustics software [32] which combines image-source and ray tracing methods to generate reverberation from a geometrical model. Using binaural rendering, their results suggest that decimating the geometry of the model had less impact on the perception of reverberation than the simplification of the frequency resolution of absorption coefficients.

In addition to plausibility and SI, perceptual dimensions such as spatial audio quality and externalization are essential for evaluating the realism of room-acoustic simulations. Spatial audio quality encompasses attributes like source distance, reverberation, and timbral coloration [see 34], while externalization describes the extent to which sounds are perceived as located outside the listener's head.

Taken together, previous studies demonstrate that considerable simplifications in acoustic geometry and simulation detail can be perceptually acceptable, particularly with regard to plausibility and SI. However, prior work has typically focused on individual environments, specific target signals, or isolated reproduction methods, limiting the ability to generalize findings across settings. Moreover, existing data are often based on separate listener groups and narrow perceptual attributes, making it difficult to assess how ALOD reductions affect different reproduction methods and room types in a unified framework.

The present study addresses this gap by systematically evaluating perceptual consequences of ALOD across three virtual environments using both headphone- and loudspeaker-based reproduction within a single group of participants. It extends previous work by including a broader range of perceptual measures, plausibility, SI, spatial audio quality, and externalization, tested under matched conditions with real-room references. This design allows for direct comparisons of ALOD effects across simulation setups and supports a more comprehensive understanding of which simplifications can be tolerated without compromising perceptual validity. Three distinct and acoustically diverse environments were investigated: BRIRs recorded in a living room coupled with a kitchen, a pub, and an underground station [34] were used as references for headphone-based auralization. Additionally, to assess the reproduction method, perceptual assessments of spatial audio quality and externalization were conducted using a three-dimensional loudspeaker array in addition to headphone reproduction. The Room Acoustics SimulatoR (RAZR) [11,13] was employed to generate synthetic BRIRs and loudspeaker reproductions.

The role of ALOD has so far often been studied using decimation of the underlying geometric model [e.g., 25,31]. While this modulates several properties of the room acoustics simulation simultaneously, such as the number and spatial direction of early reflections, the increase of echo density as well as spatial diffusion and reverberation time, the current approach aims to modulate specific, potentially important features of the simulated room impulse response in an isolated and controlled way. First, we start off with a considerably lower ALOD regarding the geometrical detail than the above studies, given that the underlying room geometry is a "proxy" rectangular (shoebox) room. With our highest ALOD model, we do not intend to model all geometrical details, however, the model has been fitted to match the frequency-dependent reverberation time and the long-term spectrum of the targeted real-world (B)RIR, important known perceptual features for comparing room acoustics. Moreover, the model has earlier been demonstrated (e.g., 18) to yield a realistic increase of echo density by incorporating sound scattering and natural-sounding (spatial) late reverberation. In order to reduce ALOD, we specifically vary selected room simulation features, such as the presence of scattering or dual-slope decays, or the inclusion of early reflections from nearby geometrical structures. Thus, ALOD was defined from a more perceptual perspective in the current study, rather than from a geometrical one defined by polygon count. The goal was to determine how ALOD can be reduced to the lowest perceptually acceptable level, depending on source stimulus and acoustic environment, starting from an already simplified and computationally efficient model. The highest ALOD condition included all available RAZR features, such as simplified effects of scattering and diffraction [18], while lower ALOD levels were created by successively omitting these components. Additionally, the effects of simulating nearby finite reflective surfaces [e.g., 35,36] and coupled volumes [13] were assessed, depending on the presence of that specific feature in the environment. This perceptually motivated definition of ALOD provides a systematic framework for assessing how different aspects of acoustic modeling complexity influence perceptual outcomes.

To motivate the choice of stimuli and perceptual tasks in this study, it is important to consider that different auditory attributes may vary in their sensitivity to changes in room modeling fidelity. Previous research has shown that stimuli with rich spatial or spectral characteristics, such as music or spatialized noise bursts, are more sensitive to rendering inaccuracies than typical speech signals [37,38]. This suggests that the detectability of reduced ALOD depends not only on the simulation method but also on the acoustic features of the stimulus. Moreover, some perceptual dimensions, particularly those related to spatial quality such as externalization or spatial clarity, may place greater demands on acoustic detail than task-specific metrics like SI [23, 39]. Accordingly, we used speech, a transient pulse, and a music excerpt for selected conditions to investigate the role of the source stimulus on sensitivity to ALOD.

The study focusses on four main research topics:

(i) Effect of ALOD variation on SI. It is expected, that SI will remain largely stable across different ALOD levels, as long as early reflections and diffuse reverberation are sufficiently represented [2,4,25,40]. However, excessive reductions in ALOD may degrade SI due to insufficient representation of early reflections that improve effective signal-to-noise ratio [41].

(ii) Effect of source stimulus on ALOD. Because different stimuli emphasize distinct acoustic cues, it is expected that transient and broadband signals (e.g., pulses) will reveal rendering differences more clearly than spectrally limited or speech-like signals, which may remain perceptually stable under reduced ALOD [42,43].

(iii) Contribution of specific spatial audio quality items on differences in perceived overall sound quality. Differences could primarily be influenced by spectral coloration, reverberation characteristics, and spatial coherence [cf. 44].

(iv) Effect of playback method (headphones vs. loudspeakers). It is expected that loudspeaker playback yields higher externalization ratings than headphone playback due to additional binaural and room cues [cf. 45,46]. However, in headphone-based simulations, externalization can still be influenced by factors such as the realism of the simulated room acoustics, including the number and spatial distribution of reflections, the balance between direct sound and early reflections, and the accuracy of spatial cues rendered through Head Related Transfer Functions (HRTFs).

## 2. Methods

### 2.1. Participants

The experiment involved eight participants with normal hearing, aged between 21 and 33 years with a mean of 28.25 years and a standard deviation of 5.85 years. All participants had prior experience in psychoacoustic tests and received an hourly compensation. For SI tests, experience varied between none and multiple hours. Two of the participants can be considered as expert listeners as defined by Kreiman et al., [47], due to them either working as a scientific researcher in audiology or by heaving experience in listening tests for at least 4 years on a regular basis.

### 2.2. Acoustic Scenes

Three everyday acoustic scenes were generated using dummy head recordings from three real-world environments, as depicted in Fig. 1 and further detailed in [34].

The first scene (left-hand column in Fig. 1) occurred in a furnished Living Room (Living Room laboratory established at the University of Oldenburg). The dimensions are 4.97 m × 3.78 m × 2.71 m, corresponding to a volume of 50.91 m³ and a reverberation time (T30) of 0.54 s. The Living Room is connected to a kitchen (4.97 m × 2.00 m × 2.71 m) through an open door, with a volume of 26.94 m³ and a reverberation time of 0.66 s. In this scenario, the receiver (black circle in Fig. 1) was seated on a sofa in the Living Room, while the target was located in the kitchen (green circle in Fig. 1), oriented towards the open door between the rooms. For the SI measurement, a masking signal was positioned on the right side of the participant (red circle in Fig. 1). The direct line of sight to the sound source was obstructed, with a path length (through the door) of 5.7 m.

The second scene (middle column in Fig. 1) was set in a Pub (OLs Brauhaus in Oldenburg), which consisted of a single large room with dimensions of 17.76 m × 10.2 m × 2.9 m, re-sulting in a volume of approximately 442 m³ and a reverberation time (T30) of 0.7 s. The receiver was positioned at a table, with the target located directly opposite, 0.97 m away (see center of Fig. 1). There was a nearby chalk board located to the left of the receiver position when pointing towards the receiver. In the second row of Fig. 1 at the bottom, the chalkboard is located at the position of the black rectangle, symbolizing a support beam. For the SI measurement, a masking signal was positions on the right side of the participant (red circle in Fig. 1). Despite its large volume, the Pub had a relatively low reverberation time.

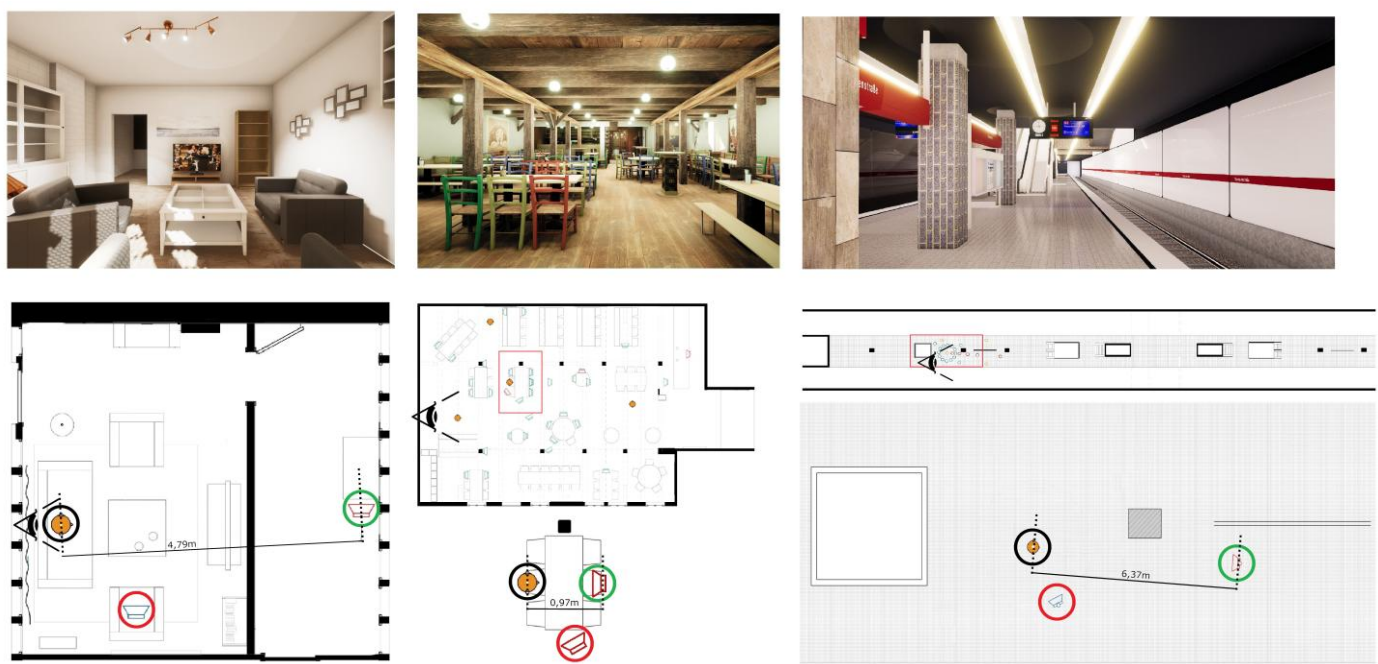

**Figure 1.** The three depicted settings, from left to right, in the Living Room, the Pub, and the Underground Station. The upper row shows the visual rendering generated by Unreal Engine. The lower row shows the floor plans, including an overview and zoomed-in areas for the Pub and Underground Station. Red rectangles highlight the zoomed-in areas in the floor plans, and an eye symbol marks the camera viewpoint from which the upper-row images were taken. The positions of the target speaker (green circle)-, masker (red circle)- and listener-position (black circle) are indicated in the bottom row.

The third scene (right-hand side of Fig. 1) was located in an Underground Station (Munich, Theresien Straße), consisting of a large, elongated space (containing the platform and two tracks) measuring approximately 120 m × 15.7 m × 4.16 m, with a total volume of about 11,000 m³ and a reverberation time (T30) of 1.6 s. Additional coupled volumes from underground tunnels and the space above the escalators contributed to a dual-slope decay in the late reverberation. The receiver was positioned near the middle of the platform, with the target signal located 6.37 m in front of the receiver. For the SI measurement, a masking signal was positions on the right side of the participant (red circle in Fig. 1).

*2.3. Room acoustic simulation*

To simulate acoustics with varying ALODs, BRIRs and loudspeaker reproductions were generated using the Room Acoustics SimulatoR (RAZR) [11,13,18]. RAZR employs a "proxy shoebox room" ISM for early reflections and a feedback delay network for diffuse reverberation using physically-based delay times. The accuracy and perceptual plausibility of RAZR have been validated in comparison to real acoustic environments [18,21,22,26]. Throughout this study, each scene was simulated with six distinct sets of features in the room acoustic simulation to vary the ALOD. This is referred to as Rendering in the following, additionally including measured BRIRs for headphone reproduction.

(1) The highest ALOD utilized RAZR with all features, including coupled rooms and a third-order ISM for early reflections. The ISM employed jittering of image source positions to avoid unrealistic, overly regular reflection patterns, and temporal smearing of early diffuse reflections to simulate scattering and multiple reflections caused by geometric irregularities and objects within the room. This configuration is referred to as RAZR. (2) In the second configuration, labeled RAZR-1st-Order, the level of detail in early reflections was reduced by lowering the ISM order from three to one. (3) The third configuration, RAZR-Simple, involved disabling specific features of the acoustic simulation for each room: For the Living Room, the coupled room simulation was simplified by performing two separate simulations for each room. The kitchen was first simulated in RAZR, with an omnidirectional receiver placed at the (closed) door. The final simulation was generated by simulating only the Living Room with an omnidirectional virtual source at the center of the (closed) door, radiating the response from the kitchen. In the Pub, the nearby reflective surfaces of the table and chalkboard were removed. In the Underground Station, the coupled volume representing the underground tunnels and escalators, responsible for the dual-slope energy decay characteristic, was omitted. (4) The fourth set, referred to as ISM, further reduced ALOD by using a straightforward implementation of a 15th-order shoebox ISM, disregarding sound scattering effects and resulting in an unnatural, completely regular pattern of specular reflections. Despite its simplicity, this type of simplified simulation is still commonly employed in room acoustics research, particularly when the goal is to isolate specific acoustic parameters or to generate computationally efficient baselines. For instance, Rathnayake and Wanniarachchi [48] used a basic ISM to explore spatial echo patterns and reverberation characteristics in 3D environments, focusing on perceptual localization effects without incorporating scattering. Similarly, Scheibler et al. [49] originally introduced Pyroomacoustics as a Python based simulation toolkit based on the image-source model for shoebox-shaped rooms. Since then, the library has been substantially extended to support more general room geometries, hybrid modeling approaches, and advanced features such as scattering and late reverberation (Pyroomacoustics documentation v0.8.6; [50]). (5) For the externalization experiment, a diotic condition, labeled Diotic, was added. For Diotic headphone tests, the left channel of the recorded signal was presented to both ears. For loudspeaker tests, Diotic was presented through a single loud-speaker positioned in front of the participant. (6) An additional control condition was generated for the SI experiment referred as Anechoic in the following. This set of BRIRs delivers no room reflections and only takes the distance between talker and listener, using the Inverse-square law, and edge diffraction for the direct sound, as it occurs in the Living Room scene between receiver and target, into account.

For headphone reproduction, a Head Related Transfer Function (HRTF) was used, that matches the artificial head used in the respective measurement of the BRIRs in the real environment. For the simulation of the Living Room, the HRTF of the database of Braren & Fels [51] was used. The database of Grimm et al. [52] was used for the Pub, and the database of Hladek & Seeber [53] was used for the Underground Station. Besides the HRTF, also the matching source directivity was used for each environment. For the Living Room and Pub, a measurement of a Genelec 8030 loudspeaker was conducted, for which the loudspeaker was placed in a rotating support in an anechoic chamber. Impulse responses were measured on a spherical grid at a distance of 2.85 m to the center of the loudspeaker, using an omnidirectional microphone. The azimuth resolution was 6°, and elevation was sampled at a step size of 3° around 0° elevation and sampled more sparsely toward the poles, for a total of 1680 directions. One hemisphere was measured and mirrored for the remaining azimuth angles. For the Underground Station the source directivity of a BM6A MK-II loudspeaker was used [53]. After generating the BRIRs in RAZR, a post-processing step was conducted in order to match the average spectrum of the simulated impulse responses to the measured ones in the frequency domain. This adjustment ensured that, within a range of 100 Hz to 16000 Hz, the average deviation between the simulated and the measured response was less than 0.5 dB across all three environments and for both the target and the masker position, based on third-octave smoothed magnitude spectra.

For loudspeaker reproduction, the simulated impulse responses were rendered over the 86-loudspekaer array (see Apparatus and Procedure) using vector-base amplitude panning (VBAP; [54]) according to the directional incidence of each component. The same spectral post-processing as for the BRIR generation was applied, using a single correction filter derived from the comparison of measured and simulated reference impulse response and applied identically to all 86 channels.

*2.4. Stimuli*

The simulated (B)RIRs and for headphone rendering, additionally the recorded BRIRs from the real environments [34] (hereafter referred to as "Measured") were convolved with one of four source stimuli. The first stimulus was speech material from the Oldenburger Satz Test (OLSA [55]), a matrix speech test composed of a large set of grammatically correct but semantically unpredictable sentences, each constructed from a total of 50 words with ten alternatives for each word type (name-verb-numeral-adjective-object). These sentences were spoken by a male speaker. For the plausibility test, eight random sentences were selected, while a single sentence was used for the evaluation of overall difference, the spatial audio quality items and externalization to allow direct comparison between the BRIR sets using the identical target signal.

The second target stimulus was a deterministic transient signal, referred to as pink pulse. This 500-ms stimulus (sampling rate of 44.1 kHz) was generated in the frequency domain by defining a pink spectrum between 50 Hz and the Nyquist frequency with minimum phase and transforming it to the time domain, following the approach described by Kirsch et al. [56]. The signal decayed from 0 dBFS to -60 dBFS within 36 ms. For the plausibility test, the spectrum of the pink pulse was modified across 10 octave bands between 31 Hz and 16 kHz. Eight variations were generated by randomly assigning each band to one of three possible level conditions, +6 dB, -6 dB or 0 dB relative to the unmodified pink spectrum, resulting in pulses with distinct tonal qualities ranging from "knocking on wood" to a "bursting balloon." For the assessment of the overall difference and externalization, the original pink-colored pulse without spectral modification was used.

For the SI experiment, a male-transformed version [57] of the International Speech Test Signal (ISTS, [58]) served as a masker. The ISTS comprises nonsensical speech synthesized from recordings of six different speakers in various languages. Lastly, a 2-second sample of an electric bass guitar was used in the overall difference spatial audio quality test.

*2.5. Apparatus and Procedure*

For the headphone experiments, participants were seated in a soundproof booth with double walls, equipped with Sennheiser HD 650 headphones connected to an RME Fireface UCX audio interface running at a sampling rate of 44.1 kHz. All listening tests were conducted using Matlab. A computer monitor was placed in front of the participants, who provided their responses using a computer mouse and keyboard. For the loudspeaker-based experiments (overall difference and externalization tasks), measurements were conducted in the VR lab at the University of Oldenburg, an anechoic chamber containing a three-dimensional array of 86 Genelec 8030 loudspeakers [see, e.g., 60]. Participants were seated at the center of the loudspeaker array, with the main horizontal ring (radius 2.4 m, height 1.8 m) consisting of 48 loudspeakers. Two additional horizontal rings, each with twelve loudspeakers, were positioned at azimuth angles of -30° and 30°, and two more rings with six loudspeakers each were positioned at azimuth angles of -60° and 60°. Two loudspeakers were also placed at -90° and 90° azimuth. Participants were seated on a chair on a platform at a height of 0.5 m so that the height of the participant's ear canal was approximately aligned with the loudspeakers in the main horizontal ring. A tablet computer was positioned in front of the participant, who used the touchscreen to provide responses. To calibrate the loudspeaker array and control the sound field at the listener's position, a G.R.A.S. 46-DP-1 1/8" LEMO Pressure Standard microphone was placed at head height in the center of the array. Delay and broadband levels for each individual loudspeaker were adjusted based on sweep measurements [cf. 60]. The sweep, which had a duration of 3.2 seconds, spanned a frequency range from 100 Hz to 22.050 Hz, and an exponential sine sweep was employed to ensure even energy distribution across octaves.

In the plausibility experiment, a single stimulus, either real (based on recorded BRIRs) or simulated, was presented through headphones. Participants were then asked, "Was the stimulus real or simulated?" This constituted a two-alternative forced-choice test. The plausibility test involved six measurements, one for each scene (Living Room, Pub, Underground Station) and for the two types of simulated impulse responses used in the test (RAZR, ISM). For each measurement, 16 target signals, comprising 8 sentences and 8 pulses (see section II.D Stimuli), were randomly presented three times, resulting in 48 presentations per measurement. Each measurement was performed twice, as both a test and retest. Before the measurements, participants underwent a brief familiarization session where they listened to example stimuli. These examples used BRIRs from different positions within the environments then those later used in the test. Both, a pulse and a sentence from the speech material were presented during familiarization. The "dry" source stimuli were initially presented without any BRIR applied, followed by the same stimuli convolved with the measured and ISM BRIRs. During the familiarization, participants were informed which signals used a measured BRIR and which were simulated.

This experiment was conducted in Matlab with the Alternative Forced Choice package (AFC [61]).

For measuring SI, the Oldenburger Satztest (OLSA [55]; see section II D. Stimuli), was conducted with AFC [61] using the closed version of this test. In this version the participants were presented a matrix of words on their screen after listening to the sentence. In this matrix all 10 response alternatives for each word type were presented and the participants task was to click on each word they understood. A spatially separated masker (ISTS) was positioned towards the right side of the listener with a close distance of about 1 meter in each environment (see red circles in Fig. 1). The masker was presented with a level of 65 dB A whereas the target sentence varied in level after each sentence, starting with a level of 70 dB A. For each test, the signal-to-noise ratio was determined at which the participant could recognize 50 % of the speech material (speech recognition threshold, SRT). Lists of twenty sentences were used for the BRIR sets Measured, RAZR, ISM and Anechoic.

For the evaluation on spatial audio quality, eight items out of the Spatial Audio Quality Inventory (SAQI) [33] were chosen, which are listed in Table 1. To obtain ratings on the spatial audio quality items, a procedure, similar to the multi-stimulus test with hidden reference and anchor (MUSHRA; [62]) was used. The measurements were performed once with headphones and once within the 86-loudspeaker array. Listeners rated various stimuli relative to a reference with regard to the respective SAQI-item using sliders on the computer screen and were able to listen to the stimuli repeatedly and sort the stimuli according to their ratings. The sliders always displayed a continuous range from 0 to 100. For the overall difference item, the slider's default position was set to 0, whereas for all other items it was set to 50. In both cases, the default position reflected "no difference to the reference". The meanings of the maximum and minimum score are listed in Table.1. A score of 50 represents no difference between the signal and the reference.

The test signals were convolved with different BRIRs: Measured (only available for headphone presentation), RAZR, ISM, RAZR-Simple and RAZR-1st-Order. RAZR served as the (hidden) reference for all Quality Items, whereas ISM effectively took the role of the anchor. For the overall difference, a single sentence of the OLSA test, the deterministic pink pulse (with no spectral variation applied) and a short sample of a played electric bass guitar were used as target. For the other spatial audio quality items, only a pink shaped pulse was used. This experiment was measured twice for each participant in a test and a retest.

**Table 1.** Spatial audio quality items used in this study, as described in Lindau et al., [33]. The terms in the middle row denote the meaning of a score of 0 whereas the terms in the right row denote the meaning of a score of 100, related to a comparison between the signal to be evaluated and the reference signal.

| *Spatial audio quality item* | *Score of 0* | *Score of 100* |
| --- | --- | --- |
| Overall difference | None | Very large |
| Distance | Closer | More distant |
| Tone color | Darker | Brighter |
| Horizontal direction | Shifted clockwise | Shifted anticlockwise |
| Metallic tone color | More pronounced | Less pronounced |
| Reverberation | More | Less |
| Envelopment | More pronounced | Less pronounced |
| Source width | Wider | Less wide |

The last experiment evaluated the perceived externalization. Again, a multi-stimulus test was used, similar to the procedure for the measurement of the overall difference, however, without a (hidden) reference. On a scale of 0 – 100, participants rated the perceived externalization with four verbal descriptors. 100 was labeled "Clearly outside the head", 66 represented "Close to the head", 33 represented "Between the ears" and 0 "Central in the head". The subjects could listen to each signal repeatedly. For this experiment, Measured, RAZR, ISM and Diotic were used. Similar to the test for the overall difference, this experiment was separated into six different tests, consisting of two different types of target signal (speech and pink pulse) and three different environments (Living Room, Pub, and Underground Station). Again, a test and a retest were conducted. Before the measurement with headphones, a short familiarization was performed. For this, the same environments were used as for the measurement, but with different positions for source and receiver. Stimuli generated with the measured BRIR were presented either diotically or binaurally to demonstrate the difference between internally and externally sounding signals.

For plausibility, spatial audio quality and externalization, the consistency of the subject responses was examined. The participant's responses for the test and for the retest were separately averaged and the test-retest correlation was calculated. A minimum correlation coefficient of 0.7 was required and participants who did not reach this value were measured a second time, where they all reached the minimum value. For the SI experiment a mean test-retest difference of maximum 1.5 dB over all conditions was required in order to be considered. However, this did not lead to re-measurements or dropped out data.

*2.6. Statistical analysis*

For all measurements, plausibility, SI, overall difference, spatial audio quality, and externalization, a repeated-measures analysis of variance (rm-ANOVA) was performed using IBM SPSS. Greenhouse–Geisser correction was applied if sphericity was violated. Post-hoc pairwise comparisons used Bonferroni correction. For the overall difference ratings, two rm-ANOVAs were conducted: First, a four-way rm-ANOVA was run with the within-subject factors Stimulus (Speech, Pulse, Bass), Environment (Living Room, Pub, Underground Station), Presentation Mode (Headphones, Loudspeaker), and Rendering (RAZR, ISM, RAZR-Simple, RAZR-1st-Order). Given that measured BRIRs were only available for headphone presentation, a second three-way rm-ANOVA was performed on the headphone data only, including the factors Stimulus, Environment, and Rendering (with the additional level Measured). This additional analysis was specifically used to compare the Measured and simulated Renderings. Likewise, for the spatial audio quality two rm-ANOVAs were conducted, but only a single Stimulus (the pink pulse) was tested. Accordingly, the first analysis included the within-subject factors Environment, Presentation Mode, and Rendering (RAZR, ISM, RAZR-Simple, RAZR-1st-Order). The second analysis was performed on the headphone data only, including Environment and Rendering, again to allow direct comparison between Measured and the simulations. Unless stated otherwise, reported statistics refer to the four-way ANOVA. A summary of all significant main effects and first-order interactions is provided in Table 2 in the Appendix.

# 3. Results

*3.1. Plausibility*

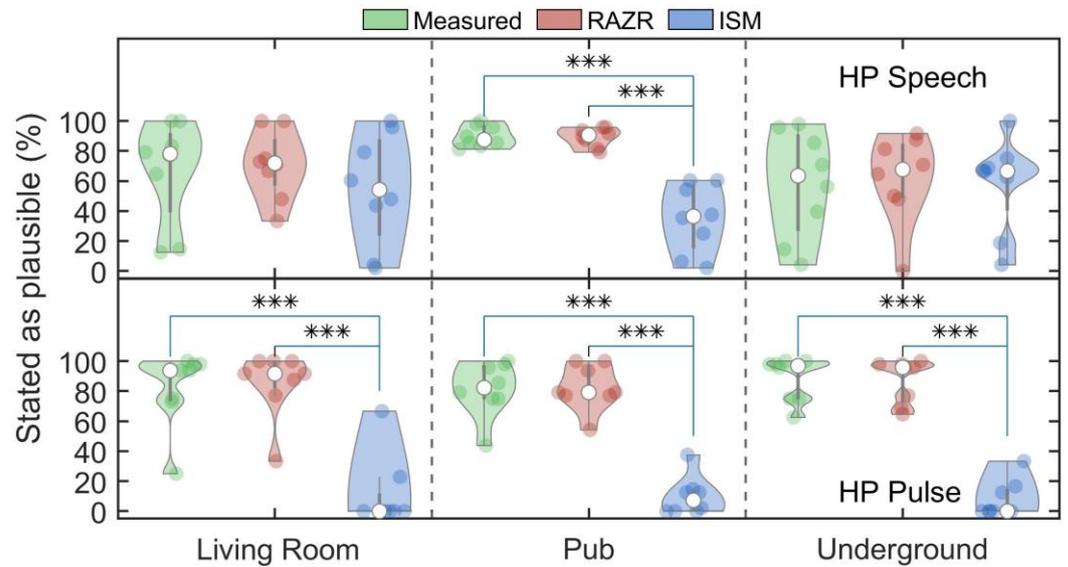

Fig. 2. Percentage of headphone auralization rated plausible for speech (top panel) and pulse (bottom panel). Green: Measured; red: RAZR; blue: ISM. Asterisks denote significant pairwise differences within each Environment and Stimulus, using post-hoc comparisons following the rm-ANOVA . *: p < 0.5 **: p < 0.01 ***: p < 0.001

The results of the plausibility test are shown in Fig. 2. The average test-retest correlation across the ten participants was 0.882 and at least above 0.8 for each participant. The top panel shows results for speech and the bottom panel for the pulse. The color of the violin plots indicates the Rendering condition. Here and in the following figures, green shows results for Measured, red shows RAZR, and blue shows ISM (from left to right). The open symbols and vertical bars indicate the median and the 25 % and 75 % percentiles. Asterisks indicate significant pairwise differences within each Environment and Stimulus, obtained from post-hoc comparisons conducted following a three-way rm-ANOVA [Stimulus (Speech, Pulse) × Environment (Living Room, Pub, Underground Station) × Rendering (Measured, RAZR, ISM)].

Pairwise comparisons revealed that in the Pub Environment, ISM was significantly less often perceived as plausible than Measured [mean difference (md) = 48.31 %; p = 0.004] and RAZR (md = 50.13 %; p < 0.001) as seen in the second row of Fig. 2, when comparing ISM (blue) with Measured (green) and RAZR (red). This was the case for both stimuli, Speech and Pulse. No significant differences between ISM and Measured and between ISM and RAZR were found when the speech material was used in the Living Room (top row, first column in Fig. 2) and in the Underground Station (top row, third column in Fig. 2).

Overall, the ANOVA revealed a significant main effect of Rendering [$F(1.034, 7.235) = 34.08$ $p < 0.001$ $\eta^2 = 0.830$] but no main effect of Stimulus or Environment. A significant interaction between Stimulus and Rendering was found. Regarding the source of this interaction, post-hoc tests showed that a significant difference between speech and pulse only occurred for ISM, as observed by comparing the blue violin plots of the top row with the blue violin plots of the bottom row. Despite this, no significant differences on plausibility were found for Stimulus or Environment.

## 3.2. Speech Intelligibility

Figure 3 shows the mean SRT for the three environments obtained with headphone presentation. The color coding follows the convention used in Fig. 2. In addition, the orange plots represent the results for Anechoic. Asterisks indicate significant pairwise dif-ferences between Renderings within each Environment, obtained from post-hoc comparisons following a two-way rm-ANOVA [Environment (Living Room, Pub, Underground Station) × Rendering (Measured, RAZR, ISM, Anechoic)]. In the Living Room, SRTs for Measured were significantly higher than for RAZR, while RAZR and ISM did not differ. Anechoic remained significantly lower than all other renderings. In the Underground Station, Measured differed significantly from ISM, while RAZR and ISM did not differ.

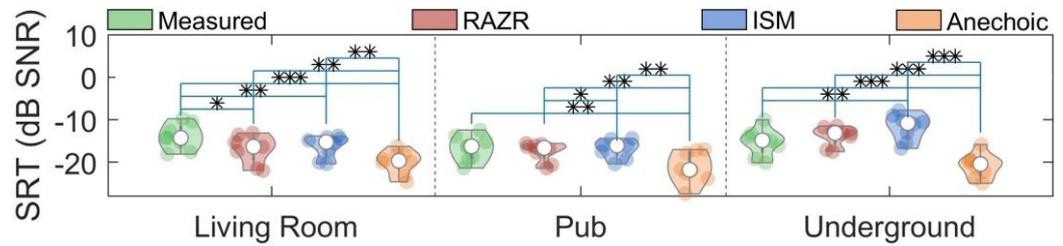

Fig. 3. Results for SI with a co-located masker, measured with headphones. Green: Measured; red: RAZR; blue: ISM; orange: Anechoic. Asterisks indicate significant differences within each environment using post-hoc comparisons following the rm-ANOVA *: $p < 0.5$ **: $p < 0.01$ ***: $p < 0.001$.

Overall, the rm-ANOVA revealed a significant main effect of Rendering on SI [$F(3, 21) = 146.48$ .; $p < 0.001$; $\eta^2 = 0.954$], with post-hoc pairwise comparisons yielding the same significant differences as indicated for the Pub in Fig, 3: RAZR (red) and ISM (blue) did not significantly differ from Measured (green), indicating that both simulated renderings reproduce SI close to the real-room reference. Anechoic (orange), by contrast, yielded significantly lower SRTs than all other renderings (md = 5.31 dB vs. Measured, $p < 0.001$; md = 4.55 dB vs. RAZR, $p < 0.001$; md = 5.72 dB vs. ISM, $p < 0.001$. Additionally, a significant difference between RAZR and ISM was found (md = 1.17 dB, $p = 0.005$) with RAZR providing the lower SRTs.

A main effect of Environment was also found [$F(1.206, 8.442) = 54.08$; $p < 0.001$; $\eta^2 = 0.885$], with generally lower SRTs in the Pub, followed by the Living Room and the Underground Station. The ANOVA also revealed a significant Rendering × Environment interaction [$F(2.36,16.54) = 12.71$ .; $p < 0.001$; $\eta^2 = 0.645$], indicating that the Rendering differences varied somewhat across Environments as shown by the asterisks. However, this interaction does not reflect an effect of Environment, but rather differences in how the renderings performed within each Environment.

## 3.3. Overall Difference

Figure 4 shows the mean ratings of overall difference for each Rendering condition, separated by Presentation Mode (upper three rows for Headphones, lower three for Loudspeakers) and by Stimulus (Speech, Pulse, Bass). The colors of the violin plots follow the color coding of the previous figures with two new Rendering conditions added here (from left to right), green: Measured (headphone presentation only), red: RAZR (reference), blue: ISM, black: RAZR-Simple; teal: RAZR-1st-Order. A value of 0 on the y-axis represents "no perceptible difference from the reference signal" and a value of 100 represents the rating "very large difference from the reference". The average test-retest correlation was 0.92, and results were pooled across repetitions. Black asterisks indicate significant pairwise differences from the reference Rendering (RAZR, red) within each Environment, Stimulus and Presentation Mode, obtained from post-hoc comparisons following a four-way rm-ANOVA [Stimulus (Speech, Pulse, Bass) × Environment (Living Room, Pub, Underground Station) × Presentation Mode (Headphones, Loudspeaker) × Rendering (RAZR, ISM, RAZR-Simple, RAZR-1st-Order)]. Red asterisks additionally indicate signif-icant differences between Measured and RAZR, obtained from post-hoc comparisons following a three-way rm-ANOVA [Stimulus (Speech, Pulse, Bass) × Environment (Living Room, Pub, Underground Station) × Rendering (RAZR, ISM, RAZR-Simple, RAZR-1st-Order)], covering the full factorial design for headphone presentation.

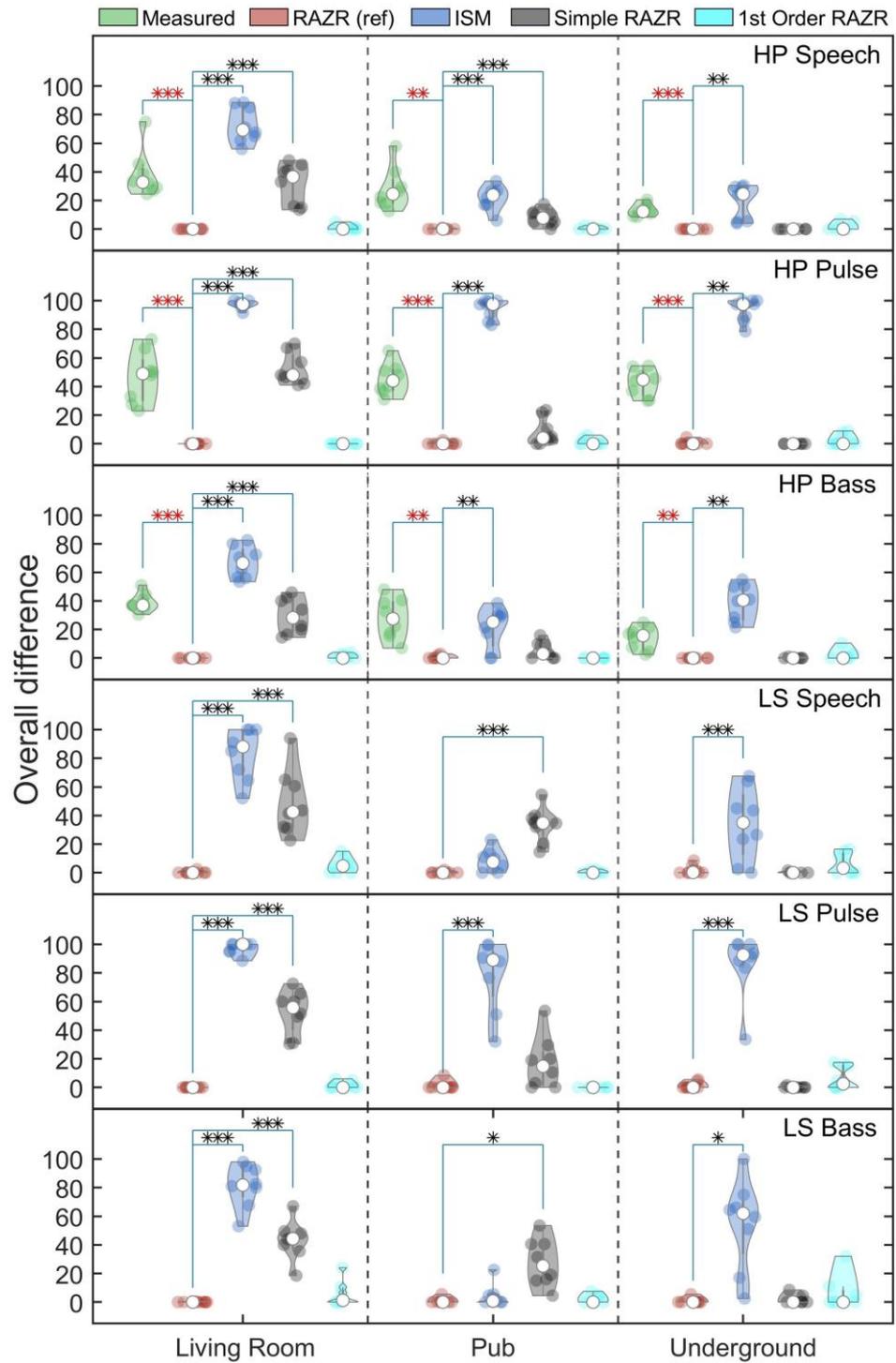

Fig. 4. Results for the overall difference rating for headphones (upper three panels showing speech, pulse and bass as target signal; HP) and loudspeaker (lower three panels; LS). Green: Measured; red: RAZR; blue: ISM; black: RAZR-Simple; teal: RAZR-1st-Order. Asterisks indicate significant differences to the reference signal within each Environment, Stimulus, and Presentation Mode using post-hoc comparisons following a three-way rm-ANOVA (red asterisks) and a four-way rm-ANOVA (black asterisks) *: $p < 0.5$ **: $p < 0.01$ ***: $p < 0.001$.

For the factor Rendering, post-hoc comparisons were conducted only relative to the reference condition (RAZR), as this rendering was available for both presentation modes and served as the baseline for all perceptual evaluations. In general, the hidden reference (RAZR) was consistently detected and rated with no difference, indicating reliability of the results. RAZR and RAZR-1st-Order (teal) were rated similar across all conditions, while ISM (blue) and Measured (green) produced the largest perceptual differences from the reference. In the Underground Station, RAZR-Simple (black) without the dual slope decay was also mostly rated with 0 difference to the reference. Confirmed by significant differences marked as black asterisks, ISM produced significantly higher difference ratings than RAZR in the Living Room and Underground Station, whereas Measured differed significantly from RAZR in all environments as indicated by the red asterisks.

The four-way rm-ANOVA revealed significant main effects of Stimulus, Environment, and Rendering, but no main effect of Presentation Mode. Specifically, Stimulus had a significant effect on the perceived overall difference [$F(2, 14) = 60.63$, $p < 0.001$, $\eta^2 = 0.896$], indicating that the rated differences depended on the type of target signal. Differences were generally rated higher for the Pulse than for Speech (md = 13.06; $p < 0.001$) or Bass (md = 12.49; $p < 0.001$) stimuli. A significant main effect of Environment was also found [$F(2, 14) = 237.90$, $p < 0.001$, $\eta^2 = 0.971$]. Post-hoc tests (not shown) revealed that larger differences occurred in the Living Room than in the Pub and Underground Station.

The main effect of Rendering was highly significant [$F(1.605, 11.235) = 470.88$, $p < 0.001$, $\eta^2 = 0.985$] and post-hoc pairwise comparisons revealed that differences were significantly larger for ISM (md = 58.04, $p < 0.001$) and RAZR-Simple (md = 19.55, $p < 0.001$) than for the reference RAZR. By contrast, RAZR and RAZR-1st-Order differed only slightly (md = 0.85, $p = 0.039$). This pattern is clearly visible in Fig. 4, where both Renderings show near-zero median differences across all stimuli. Across environments and stimuli, RAZR and RAZR-1st-Order were rated most similar, while ISM and RAZR-Simple produced substantially higher difference ratings.

Results from the additional three-way rm-ANOVA (headphone data only) showed a significant main effect of rendering and the related post-hoc tests confirmed that the Measured condition was consistently rated as significantly different from the RAZR reference across all environments as indicated by the red asterisks in Fig. 4. No significant main effect of Presentation Mode was found [$F(1, 7) = 2.32$, $p = 0.172$, $\eta^2 = 0.249$], indicating that overall difference ratings were similar between headphone and loudspeaker presentation. A Stimulus × Environment interaction was found [$F(4, 28) = 9.99$, $p < 0.001$, $\eta^2 = 0.588$)]. Post-hoc tests showed that this interaction was driven by the Bass stimulus: ratings in the Living Room differed significantly from those in the Pub and the Underground Station. For Speech and Pulse, no such differences between environments occurred.

### 3.4. Spatial Audio Quality Items

Figure 5 shows the results for headphone (A, left) and for loudspeaker (B, right) presentation. Each row in both main panels represents one spatial audio quality item and each column an environment. The horizontal
line in each plot marks a score of 50, corresponding to "no perceptual difference between the test signal and the reference." The color coding follows the convention of Fig. 4: green = Measured (headphones only), red = RAZR (reference), blue = ISM, black =

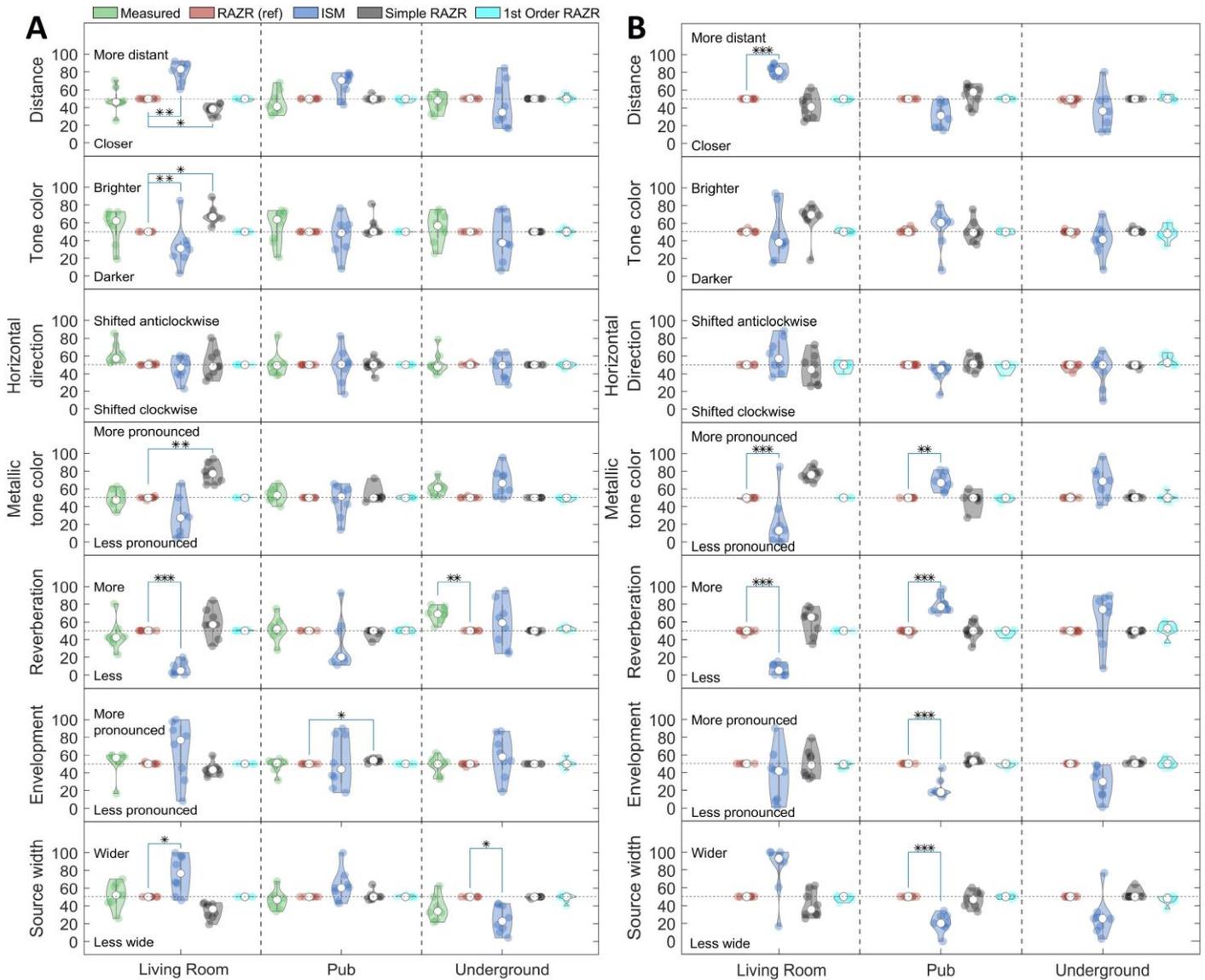

Fig. 5. Results for the spatial audio quality rating for headphones (A, left) and loudspeakers (B, right). Green: Measured (headphones only); Red: RAZR; blue: ISM; black: RAZR-Simple; teal: RAZR-1st-Order. Asterisks indicate significant differences to the reference signal within each Environment, Stimulus, and Presentation Mode using post-hoc comparisons. *: p < 0.5 **: p < 0.01 ***: p < 0.001

RAZR-Simple, teal = RAZR-1st-Order. Asterisks indicate significant pairwise differences from the reference (RAZR) within each environment, obtained from post-hoc comparisons following a three-way rm-ANOVA [Environment (Living Room, Pub, Underground Station) × Presentation Mode (Headphones, Loudspeaker) × Rendering (RAZR, ISM, RAZR-Simple, RAZR-1st-Order)] performed for each spatial audio quality attribute. For the factor Rendering, post-hoc comparisons were performed only relative to RAZR, as it served as the common reference across presentation modes. Additional two-way rm-ANOVAs [Environment (Living Room, Pub, Underground Station) × Rendering (Measured, RAZR, ISM, RAZR-Simple, RAZR-1st-Order)] for the headphone data (Fig. 5A) revealed no significant differences between the Measured condition and the RAZR reference for any of the sound-quality items.

A first inspection of Fig. 5 shows that large perceptual differences between renderings occurred only occasionally. When present, these differences were most pronounced for ISM (blue) and, to a lesser extent, for RAZR-Simple (black) in the Living Room. The data also show considerable inter-individual variability, resulting in relatively few statistically significant differences overall, as reflected by the sparse asterisks in each panel. No major differences could be found between the two Presentation Modes Headphones and Loudspeaker. For headphones (Fig. 5A), Measured (green) showed small deviations from the RAZR reference (red); however, these differences often occurred in both positive and negative directions across participants and environments, resulting in mean ratings that remained centered around the reference, between 40 and 60. Consequently, no significant differences between Measured and RAZR were observed, except for Source Width in the Underground Station. For tone color, a small but consistent (yet non-significant) difference is visible between Measured and RAZR for all environments.

The following description summarizes the significant main effects, 1st-order interactions, and post-hoc results obtained from the three-way rm-ANOVAs:

Effect of Environment. The three-way rm-ANOVAs revealed significant main effects of Environment for the items Distance ($\eta^2 = 0.69$), Reverberation ($\eta^2 = 0.81$), Source Width ($\eta^2 = 0.67$), Tone Color ($\eta^2 = 0.40$), and Metallic Tone Color ($\eta^2 = 0.40$). In general, perceived differences from the reference in Distance and Reverberation were rated higher in the Living Room than in the Pub and Underground Station.

For Reverberation, significant interactions were observed with both Presentation Mode and Rendering. A significant Environment × Presentation Mode interaction showed that, for headphone presentation, perceived differences from the reference were significantly larger in the Living Room than in the Underground Station, whereas for loudspeaker presentation, the Living Room differed from both the Pub and the Underground Station. This indicates that environment-dependent deviations in perceived reverberation from the reference were more pronounced under loudspeaker playback.

Furthermore, a significant Environment × Rendering interaction indicated that these Environment-dependent variations occurred primarily for the ISM rendering, whereas the other renderings showed more consistent ratings across environments. This pattern suggests that the reverberant character reproduced by ISM was particularly sensitive to the acoustic context, resulting in stronger perceived deviations from the reference in the Living Room.

Several other higher-order interactions between Environment and other factors (e.g., Rendering) reached statistical significance; however, post-hoc comparisons revealed no systematic or interpretable trends beyond the main effects. Occasional differences appeared for individual renderings within single environments (e.g., ISM or RAZR-Simple in the Living Room), but these effects were inconsistent across items and did not alter the overall pattern of results. Therefore, only the main effects and interactions that revealed coherent, perceptually meaningful differences are reported in detail.

Effect of Rendering. The three-way rm-ANOVAs revealed significant main effects of Rendering for the items Distance ($\eta^2 = 0.55$), Metallic Tone Color ($\eta^2 = 0.43$), and Reverberation ($\eta^2 = 0.43$), indicating that the perceived deviation from the reference depended on the rendering method. In general, ISM and RAZR-Simple led to the largest perceptual deviations from the reference (i.e., ratings that were shifted furthest away from the "no difference" point at 50), whereas RAZR and RAZR-1st-Order produced ratings closest to 50 and were therefore most similar to the reference.

For Distance, post-hoc comparisons showed a significant difference only between RAZR-Simple and RAZR-1st-Order. No other pairwise comparisons between renderings reached significance. As seen in the first column, first row of Fig. 5A, this effect was mainly driven by the Living Room condition, where RAZR-Simple produced larger deviations from the reference than RAZR-1st-Order. In the other environments, the two renderings were more similar. No consistent differences were observed between the remaining renderings.

For Metallic Tone Color, post-hoc tests revealed significant differences between RAZR and RAZR-Simple, and between RAZR-Simple and RAZR-1st-Order. No other rendering pairs differed significantly. In the Living Room (Fig. 5A, fourth row, first column), RAZR-Simple produced a perceptual shift in metallic coloration relative to both the full RAZR and the RAZR-1st-Order conditions, suggesting that this attribute was particularly sensitive to the simplified coupled-room configuration used in RAZR-Simple.

For Reverberation, although the main effect of Rendering was statistically significant, none of the pairwise post-hoc comparisons between renderings reached significance. Thus, the overall effect of Rendering on perceived reverberation differences from the reference was small and not attributable to a specific pair of renderings. Several significant interaction effects were observed between the experimental factors, most prominently involving Rendering and Environment.

For the items Distance, Metallic Tone Color, and Reverberation, the Rendering × Environment interactions followed a consistent pattern. Significant rendering differences occurred mainly in the Living Room and were driven by the ISM and RAZR-Simple renderings, both of which were rated as perceptually more different from the reference than the other renderings. The remaining renderings showed more stable ratings across environments and remained perceptually closer to the reference.

Two notable exceptions were observed. For Source Width, significant rendering differences appeared only in the Underground Station, where ISM again differed from all other renderings. For Envelopment by Reverberation, a Rendering × Presentation Mode interaction indicated that significant differences between RAZR and ISM occurred only under Loudspeaker presentation, but not with Headphones. Overall, these findings indicate that rendering-related perceptual deviations were most pronounced for ISM and, to a lesser extent, RAZR-Simple, particularly in the Living Room and Underground Station.

Effect of presentation mode. A significant main effect of Presentation Mode was found for Envelopment by Reverberation ($\eta^2 = 0.46$). Ratings were, on average, about seven points lower under Loudspeaker presentation than under Headphone presentation, indicating a reduced sense of envelopment when reproduced over loudspeakers.

A significant Rendering × Presentation Mode interaction was also observed for Envelopment by Reverberation. Post-hoc comparisons showed that this difference between presentation modes occurred only for the ISM rendering, whereas all other renderings yielded comparable ratings for headphones and loudspeakers. In addition, a significant Environment × Presentation Mode interaction was found for the item Reverberation. Here, the Pub environment showed a presentation-dependent difference: under Headphone presentation, reverberation was rated slightly lower (−4.2 points relative to the reference), whereas under Loudspeaker presentation it was rated higher (+6.7 points). No such differences between presentation modes were observed in the other environments.

Overall, these findings suggest that the influence of presentation mode on the spatial audio quality ratings was limited and primarily affected perceived Envelopment, particularly for the ISM rendering and in the Pub environment.

*3.5. Externalization*

Figure 6 shows the mean externalization ratings for each Rendering condition, separated by Presentation Mode (upper two rows: Headphones; lower two rows: Loudspeakers) and by Stimulus (Speech, Pulse). The color coding follows the convention of the previous figures: green = Measured (headphones only), red = RAZR, blue = ISM, yellow =

Diotic. The average test–retest correlation was 0.84, and results were pooled across repetitions. Asterisks indicate significant pairwise differences between renderings within each Environment, Stimulus, and Presentation Mode, obtained from post-hoc comparisons following a four-way rm-ANOVA [Stimulus (Speech, Pulse) × Environment (Living Room, Pub, Underground Station) × Presentation Mode (Headphones, Loudspeaker) × Rendering (RAZR, ISM, Diotic)]. An additional two-way rm-ANOVA [Environment (Living Room, Pub, Underground Station) × Rendering (Measured, RAZR, ISM, Diotic)] was conducted in order to analyze differences between Measured and the simulation for the headphone data.

Across both Stimuli, ratings for RAZR and Measured produced the highest externalization ratings, followed by ISM. For headphone measurements, Diotic was the least externalized Simulation, while it was among the most externalized simulations for the Loudspeaker presentations. For Speech, externalization was generally weak under headphone presentation across all renderings, suggesting that listeners found speech particularly prone to in-head localization in this playback condition, whereas the Diotic rendering yielded the poorest externalization overall, as was expected.

In the following, only main results and first order interactions from the four-way ANOVA are reported. The three-way ANOVA did not reveal any additional relevant findings. The four-way rm-ANOVA revealed significant main effects of Presentation Mode $[F(1,7) = 9.39 ; p = 0.018 ; \eta^2 = 0.57]$ and Rendering $[F(2,14) = 8.04 ; p = 0.005 ; \eta^2 = 0.54]$. Overall, sounds were rated as more externalized under loudspeaker presentation than under headphone presentation, indicating that externalization benefited from the additional spatial cues available during loudspeaker playback.

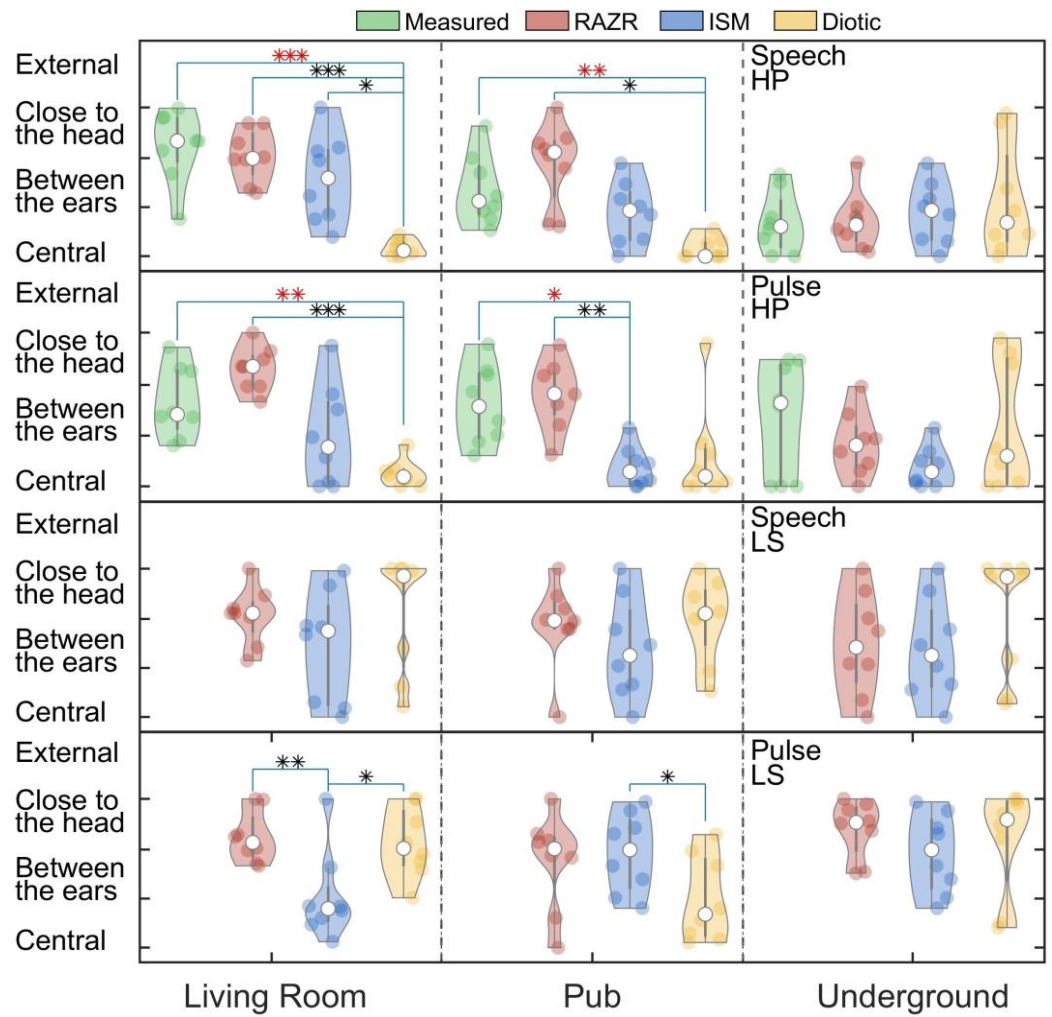

Fig. 6. Results for the externalization ratings for headphones (upper two panels; HP) ) and loudspeaker (lower two panels; LS) in the same style as in Fig. 3. Green violin plots: Measured; red plots: RAZR; blue plot: ISM; yellow plots: Diotic. Asterisks indicate significant differences between Renderings within each Environment, Stimulus, and Presentation Mode using post-hoc comparisons following a three-way rm-ANOVA (red askerisks) for comparisons with Measured and a four-way rm-ANOVA (black askerisks) *: p < 0.5 **: p < 0.01 ***: p < 0.001.

Post-hoc comparisons confirmed that RAZR was rated significantly more externalized than ISM (p < 0.05), whereas the difference between RAZR and Diotic did not reach significance.

A significant Environment × Rendering interaction in the four-way rm-ANOVA was observed. In the Living Room, ISM was rated significantly less externalized than RAZR, while in the Pub and Underground Station, no significant differences between renderings were found. This pattern suggests that the rendering-dependent differences in perceived externalization were most pronounced in less reverberant acoustic contexts. A significant Presentation Mode × Rendering interaction was also found in the four-way rm-ANOVA. Post-hoc tests revealed that this interaction was driven solely by the Diotic condition: for headphone presentation, Diotic signals were rated strongly internalized (mean = 18.8), whereas for loudspeaker presentation, the same condition yielded substantially higher externalization ratings (mean = 63.1; md = 44.4, p < 0.001). For RAZR and ISM, no significant differences between headphone and loudspeaker presentation were observed.

Taken together, these results show that perceived externalization was strongly influenced by both presentation mode and rendering method. Externalization was consistently highest for RAZR, lowest for Diotic, and moderately reduced for ISM, with the most pronounced rendering effects occurring under headphone presentation and in the Living Room environment.

## 4. Discussion

### A. Role of ALOD across differrent perceptual measures and stimuli

The perceptual influence of the ALOD varied considerably across perceptual measures, Environments, Stimulus types, and Rendering conditions. For the speech stimulus, plausibility judgments were relatively robust to reductions in ALOD as long as essential acoustic cues such as the overall reverberation time and the basic pattern of early reflections were maintained. In both the Living Room and Underground Station, even the simplest ISM rendering achieved plausibility ratings close to those obtained with the measured BRIRs. This suggests that for speech, which contains limited spectral content and temporal modulations lacking strong transients, perceptual plausibility does not rely on a highly detailed representation of the reverberant sound field as long as the general acoustic characteristics of the Environment are preserved.

SI exhibited a similar pattern of robustness. Across all Environments, SRTs remained stable for most Renderings and showed no significant degradation compared to the measured reference. This aligns with established evidence that the course pattern of early-arriving reflections and the direct-to-reverberant energy ratio, both of which are maintained in all renderings, are dominant factors affecting speech perception in rooms [40,42]. Only in the Underground Station did ISM lead to significantly higher SRTs than for Measured. This is likely related to an unrealistically sparse pattern of reflections obtained in the shoebox model of the room in connection with an overly slow increase of echo density. This property of the ISM has been shown for a large room volume in Ewert et al. [18] using the normalized echo density (NED; [63]), demonstrating that a slow buildup of echo density reduces the temporal and spatial continuity of the reverberant field and thereby impairs SI in acoustically complex spaces.

The sensitivity to ALOD increased markedly for the Pulse and Bass stimuli, where transient onsets and broadband spectral content accentuated differences between Renderings. Plausibility ratings dropped sharply for ISM in all Environments, and Overall Difference ratings showed the largest deviations from the reference for ISM and RAZR-Simple. These findings demonstrate that renderings with reduced ALOD, such as fewer early reflections, missing scattering effects, or overly regular reflection patterns, fail to reproduce the fine temporal and spectral cues that are perceptually relevant for transient or broadband stimuli.

For the spatial audio quality items, which were evaluated using only the "most critical" Pulse stimulus, the effects of reduced ALOD were particularly evident in the Living Room: Both ISM and RAZR-Simple were perceived as significantly different from the ref-erence, particularly for Distance, Reverberation, and Metallic Tone Color. These deviations point to inaccuracies in the representation of coupled-room effects and the spatial distribution of early reflections. In RAZR-Simple, the simplified coupled-room configuration did not yield a perceptually equivalent reproduction of the reference, underlining that modeling simplifications in geometrically complex Environments directly impair perceived spatial audio quality. Although the Overall Difference results in Figure 4 revealed significant perceptual deviations between the Measured and RAZR, these differences were not mirrored in the ratings of the individual spatial audio quality attributes. This suggests that overall difference may reflect a composite percept that integrates several of the tested spatial audio quality attributes even when their individual contribution did not reach significance, such as spectral coloration, or was differently but inconsistently rated. Alternatively, the underlying spatial audio quality differences may not have been fully captured by the here selected seven attributes. Future studies could address this by expanding the attribute set or by applying data-driven analyses, such as multidimensional scaling, to identify the dominant perceptual dimensions contributing to global judgments of simulation fidelity.

The influence of ALOD reductions in RAZR-Simple was particularly evident in the Pub, where Overall difference ratings showed larger deviations from the reference. These deviations can be attributed to the omission of two nearby reflecting surfaces, a chalkboard and a table, resulting in a perceptible change in spatial impression. Listeners reported a missing lateral reflection from approximately 30°–40° to the left of the source direction, where the chalk board is positioned. This omission was particularly salient under loudspeaker presentation, where natural binaural cues and head movements enhanced spatial resolution. The higher overall-difference ratings for RAZR-Simple under loudspeaker playback thus likely reflect the absence of these distinct early reflections.

In contrast, for the Underground Station, RAZR-Simple differed from RAZR primarily through the omission of the dual-slope decay in the late reverberation tail. Despite this physical simplification, no perceptual differences were found between the two renderings. Analysis of the normalized energy decay curves shows that the secondary slope began only at a very low energy level (below roughly −40 dB), where its contribution to perceived reverberant character is minimal. Consequently, both versions produced nearly identical decay behavior within the time-intensity region most relevant to perception. This explains why the absence of the dual-slope tail did not lead to perceptual degradation, even though the underlying acoustic model was simplified.

Externalization results provided additional evidence that variations in the ALOD influences spatial aspects of auditory perception. RAZR produced the highest externalization ratings, followed by ISM, while Diotic signals were perceived as strongly internalized. The reduced externalization for ISM suggests that missing spatial detail and inaccurate room-related cues limit the perception of external auditory space. Similar tendencies have been reported in studies linking spatial externalization to the precision of binaural and room-related cues [37]. Interestingly, across all three environments, the Diotic condition was in most cases rated as more externalized than the simulated renderings, with the exception of the Pub for the Pulse stimulus. This finding may be explained by the loudspeaker presentation setup. In the Diotic condition, a single loudspeaker positioned 2.4 m in front of the listener was used. For some participants, this clear frontal localization may have been perceived as more external than the more spatially diffuse sound fields generated by the multichannel renderings, which distribute acoustic energy across multiple loudspeakers from different directions. Such increased spatial diffuseness can reduce externalization by weakening the consistency between auditory and spatial cues [46,64].

One important difference between the current ALOD variations and earlier studies such as, [31] and [25], is that the current acoustic model was always based on a "proxy" shoe-box simplification of each room, as outlined above. Thus, even the highest ALOD, the pattern of early reflections was always approximate rather than geometrically exact. Despite this simplification, the RAZR rendering produced plausibility, SI, spatial audio quality and externalization ratings that closely matched those obtained with the measured reference, confirming that perceptual agreement can be achieved even when using simplified geometries, provided that key temporal and spectral room characteristics are accurately represented [21]. Nevertheless, the shoebox approximation becomes more critical for lower ALOD conditions, particularly for ISM, where the lack of geometric detail led to sparse and overly regular reflection patterns. Such patterns reduce echo density and temporal diffusion, impairing the continuity of the reverberant field, as discussed above and shown for large-volume spaces in [18]. This highlights that while high-ALOD shoebox simulations can reproduce perceptual realism when properly parameterized, further simplifications risk perceptual degradation due to insufficient spatial and temporal reflection density.

Taken together, these findings indicate that the perceptual importance of ALOD depends strongly on the specific perceptual measure as well as on the stimulus type. Measures associated with speech as a stimulus and speech communication in general, such as plausibility and intelligibility, are comparatively tolerant to rendering simplifications as long as the overall reverberation time, the balance between direct and reverberant energy, and the general distribution of early reflections are preserved. In contrast, broadband and transient stimuli such as the Pulse emphasized differences in reflection timing, coloration, and spatial coherence, leading to stronger perceptual deviations when ALOD was reduced [cf. 56]. For the Bass stimulus, overall difference ratings were lower across environments and renderings. Across all stimuli, higher ALOD was particularly critical for perceptual dimensions relying on fine spatial or spectral cues, such as spatial audio quality attributes and externalization. The results for RAZR-Simple further demonstrate that reducing geometric detail, particularly in coupled spaces, can lead to noticeable perceptual deviations.

**B. Role of environment**

The acoustic Environment had a pronounced influence on the sensitivity to changes in ALOD across most perceptual measures. Across perceptual measures, the Living Room consistently produced the largest deviations from the reference, while the Underground Station and the Pub generally yielded smaller or more stable differences. The Living Room was particularly challenging to simulate due to its coupled-room configuration, which introduces complex reflection patterns and modal interactions that are difficult to reproduce accurately with simplified geometric models.

For plausibility and spatial audio quality, both ISM and RAZR-Simple produced perceptually larger deviations from the reference in the Living Room than in the other Environments. The results suggest that the critical coupled-volume geometry amplified perceptual differences between Renderings, particularly for attributes related to reverberation and tonal coloration. This is consistent with earlier studies demonstrating that the perception of spatial fidelity in coupled or irregularly shaped rooms depends strongly on the correct reproduction of reflection density and decay characteristics [23,37].

In contrast, the Pub and Underground Station exhibited greater perceptual tolerance to simplified modeling for plausibility, spatial audio quality and externalization. In these Environments, the long or diffuse reverberation provided a masking effect that reduced sensitivity to errors in spatial detail. In the Underground Station, interestingly plausibility ratings for ISM remained close to the reference; however, SI was reduced due to the unrealistically sparse reflection pattern and slow buildup of echoic density characteristic of the ISM, as discussed above. These deficiencies likely disrupted the temporal continuity of the reverberation and thereby impaired speech cue transmission. Similar masking effects of diffuse late energy on spatial perception have been observed in previous studies [42,65]. These SI results for the Underground Station align well with those reported by Hladek and Seeber [53], who measured SRTs at the same position (R1 to pos. 16). Likewise, Schütze et al. [4] evaluated SI at Position S7 in the real Living Room with the same location for receiver and target source that was used in our study, and reported SRTs comparable to our measured reference. These findings support the correctness of our measurements.

The Environment also influenced the results for individual spatial audio quality attributes. Perceived Distance and Reverberation were rated most different from the reference in the Living Room, whereas Source Width and Metallic Tone Color showed smaller variations across Environments. These findings correspond to the significant Environment × Rendering interactions, which were mainly driven by the ISM rendering. The ISM produced particularly large deviations in the Living Room, indicating that its reverberation was more affected by the acoustic complexity of the environment. In contrast, the other renderings yielded more consistent rating across Environments, suggesting that the perceptual impact of ALOD reductions depends on both the simulated environment and the rendering method.

For externalization, the influence of the Environment was limited overall but became apparent in the Living Room, where ISM was perceived as significantly less externalized than RAZR, as the Environment × Rendering interaction showed. This indicates that in acoustically complex but less diffuse spaces such as the Living Room, listeners were more sensitive to rendering inaccuracies that disrupt binaural or room-related cues, which are known to support externalization [45,64]. In contrast, in the more reverberant Pub and Underground Station, late reflections masked these differences, leading to comparable externalization ratings across renderings.

A closer examination of the Underground Station data revealed that externalization ratings for speech were generally low across all renderings under headphone presentation. This finding can be explained by the room's specific acoustic characteristics: although it exhibits a long reverberation time, the overall reverberation level is comparatively low. As a result, much of the reverberant energy remains masked by the speech and therefore contributed little to externalization [42,45]. In contrast, for the Pulse stimulus, the reverberant decay was always audible, allowing late reflections to enhance spatial impression and perceived externalization. In this condition, the Measured rendering produced slightly higher externalization ratings than RAZR, likely reflecting the additional fine spatial and spectral details contained in the measured room responses.

Taken together, these findings demonstrate that the acoustic properties of the Environment shape how listeners evaluate rendering fidelity. In acoustically simple or diffuse spaces, errors in simulation detail are perceptually less relevant due to reverberant masking. However, in Environments with coupled volumes, strong directional reflections, or complex geometries, perceptual sensitivity to rendering accuracy increases considerably. Consequently, the required level of ALOD depends not only on the perceptual measure but also on the acoustic structure of the Environment being reproduced.

**C. Role of Presentation Mode**

Presentation mode exerted a differentiated influence on spatial perception across the evaluated measures. For Overall Difference, headphone and loudspeaker presentations yielded statistically similar ratings, indicating that the perceived global similarity between renderings was largely independent of playback mode. In contrast, for measures directly related to spatial hearing, such as externalization, presentation mode had a pronounced effect, with loudspeaker playback generally enhancing spatial impression. This suggests that participants were able to form consistent judgments about rendering fidelity even when the spatial information was conveyed through individualized binaural playback rather than through free-field loudspeaker reproduction.

In contrast, several spatial audio quality attributes were modestly affected by presentation mode. The most notable difference appeared for Envelopment by Reverberation, where ratings were, on average, lower under loudspeaker presentation. This reduction likely reflects the additional spatial cues and room interactions introduced by the physical playback environment, which may have altered listeners' perception of the simulated reverberant field [43,44]. For the ISM rendering, this effect was particularly pronounced, suggesting that limited spatial coherence in the simulation was perceptually masked under headphones but revealed more clearly when compared to the real acoustic context of loudspeaker playback. A smaller, environment-specific presentation mode effect also occurred in the Pub, where reverberation was rated slightly weaker under headphones and stronger under loudspeakers, possibly due to differences in the integration of simulated and real spatial cues.

The influence of presentation mode was most apparent for Externalization, which showed a strong and systematic dependence on playback condition. Externalization ratings were generally higher under loudspeaker presentation, reflecting the contribution of naturally occurring head movements and dynamic spatial cues that reinforce external perception [45,66]. Under headphone playback, where such cues were absent, listeners were more susceptible to in-head localization, especially for the Diotic rendering. This rendering produced strongly internalized images during headphone presentation but was perceived as substantially more externalized when played over loudspeakers, confirming that the availability of congruent spatial cues from the playback setup can compensate for missing binaural information. In contrast, RAZR and ISM maintained stable externalization ratings across presentation modes, indicating that their spatial renderings provided sufficient directional consistency to support externalization even under headphone reproduction.

Taken together, these findings demonstrate that presentation mode primarily affects perceptual dimensions that depend on dynamic or spatially extended cues, such as externalization and envelopment. When accurate room simulation is available, headphone reproduction can yield perceptually equivalent results for most spatial audio quality attributes, but for the perception of spatial extent and realism, loudspeaker playback continues to offer a distinct advantage.

D.Comparison to geometry decimation and outlook

Starting from the current highest ALOD, which already includes substantial geometric simplifications by using a proxy rectangular room, the results showed that particularly for speech, further reductions of ALOD were perceptually tolerable—for instance, reducing the number of early reflections or omitting scattering effects, as in the ISM. It should be noted that this lowest tested ALOD (ISM) would actually be more computationally demanding than the highest ALOD (RAZR), again underlining that the present definition of ALOD is not based on geometric or computational complexity but rather on perceptual properties of the room simulation. The inclusion of nearby reflecting surfaces in the Pub environment specifically added geometric detail to the room acoustics simulation, improving over the basic shoebox simplification for early reflections. In the current results, this addition proved perceptually important, as the omission of these reflectors in RAZR-Simple led to clearly detectable spatial differences under loudspeaker presentation.

Bridging the gap to studies that have investigated geometry decimation [e.g., 25,31], future research should extend the current approach by systematically increasing geometric detail stepwise from the shoebox approximation to assess potential benefits in conditions where the real environment cannot be meaningfully modeled by connected rectangular rooms, such as connected corridors, obstructing structures, or columns. The role of edge diffraction has already been examined in recent studies [e.g.67] and our current modeling framework (de Haas et al., 2025) has been extended to cover arbitrary geometry and edge diffraction using computationally efficient filter-based approximations [35,36,68,69].

## 6. Conclusion

Plausibility, SI, spatial audio quality, and externalization were investigated with different acoustic level of detail of the room acoustics simulation in three (real world) environments with substantially different size and (room acoustic) properties. Headphone and loudspeaker reproduction were tested, including comparison to recordings from the real environments for headphones. It was demonstrated that highly plausible simulations with externalization comparable to dummy head recordings can be achieved with considerable simplifications in the acoustic simulation: Even the most detailed ALOD applied in this study utilized a "shoebox" approximation of room geometry for early reflections. However, scattering effects were incorporated, and diffuse late reverberation was modeled using a physics-based feedback delay network. Overall, the results indicate that ALOD should be chosen in relation to the intended application: more reduced models are suitable for communication-oriented simulations using speech stimuli, whereas broadband, transient stimuli require higher fidelity of the model:

- •SI remained largely stable across all tested ALOD levels including the highly simplified shoebox image source model. For the large volume of the Underground Station, overly sparse early reflections and a too slow rise of echo density in the image source model are a possible source for the observed deviations for SI.
- •Speech stimuli proved relatively robust, producing only small perceptual differences between renderings with different ALOD, whereas the broadband, transient pulse stimulus revealed pronounced perceptual differences for simplified simulations. Plausibility and overall-difference ratings were particularly sensitive to ALOD reductions for the pulse.
- •No clear cut contribution of specific spatial audio quality attributes to the observed overall difference between simulated and measured environments was found. There was a trend, supporting that coloration differences, although not significant, in connection with perceived differences in reverberation, distance, and metallic tone color might be the most influential attributes.
- •While overall difference and spatial audio quality ratings were largely comparable for both (static) headphone and loudspeaker array reproduction, externalization was clearly improved for loudspeaker reproduction, reflecting the contribution of dynamic spatial cues caused by natural head movements and the own head-related transfer function to the perception of external space. For headphones, simulations with the highest ALOD reached comparable levels of externalization as observed for the dummy head recordings in the real room.

## 6. Patents

**Author Contributions:** Conceptualization, S.F. and S.E.; methodology, S.F. and S.E..; software, S.F.; validation, S.F..; formal analysis, S.F.; investigation, S.F. and S.E.; resources, S.E.; data curation, S.F.; writing—original draft preparation, S.F. and S.E.; writing—review and editing, S.F., S.E., S.v.d.P.

and B.S.; visualization, S.F.; supervision, S.E. and S.v.d.P.; project administration, S.E. and B.S. ; funding acquisition, S.E. and B.S.

All authors have read and agreed to the published version of the manuscript

**Funding:** This work was funded by the Deutsche Forschungsgemeinschaft (DFG, German Research Foundation) – Project-ID 352015383 – SFB 1330 C5.

Data Availability Statement: Data are available upon request.

**Acknowledgments:** The authors thank Christoph Kirsch for his support with the room acoustics simulations.

During the preparation of this manuscript, the authors used OpenAI ChatGPT for language polishing. The authors have reviewed and edited the output and take full responsibility for the content of this publication.

**Conflicts of Interest:** The authors declare no conflicts of interest. The funders had no role in the design of the study; in the collection, analyses, or interpretation of data; in the writing of the manuscript; or in the decision to publish the results.## Abbreviations

The following abbreviations are used in this manuscript:

| | |
|---|---|
| ALOD | Acoustic Level of Detail |
| AR | Augmented Reality |
| VR | Virtual Reality |
| LOD | Level of Detail |
| SI | Speech Intelligibility |
| BRIR | Binaural Room Impulse Response |
| CATT | Computer Aided Theater Technique |
| RAZR | Room Acoustics SimulatoR |
| HRTF | Head Related Transfer Function |
| ISM | Image Source Model |
| VBAP | Vector Based Amplitude Processing |
| OLSA | Oldenburger Satz Test |
| ISTS | International Speech Test Signal |
| SRT | Speech Recognition Threshold |
| MUSHRA | Multi Stimulus Test with Hidden Reference and Anchor |
| SAQI | Spatial Audio Quality Inventory |
| rm-ANOVA | Repeated Measures Analysis of Variance |

## Appendix A

**Table A1.** Results of repeated-measures ANOVAs for all perceptual measures. The left panel includes headphone and loudspeaker presentations (excluding the Measured condition), and the right panel shows headphone data including Measured. Reported are degrees of freedom ($df_M$, $df_R$), F-values, p-values, and effect sizes ($\eta^2$) for significant main effects and interactions.

| ANOVA data with loudspeaker and headphone presentation without | | | | | | ANOVA data with headphone presentation including Measured | | | | | |
|---|---|---|---|---|---|---|---|---|---|---|---|
| Factor | $df_M$ | $df_R$ | $F$ | $p$ | $\eta^2$ | Factor | $df_M$ | $df_R$ | $F$ | $p$ | $\eta^2$ |
| **Distance** | | | | | | **Distance** | | | | | |
| Environment | 2 | 14 | 15.43 | <0.001 | 0.69 | | | | | | |
| Rendering | 1.12 | 7.81 | 8.61 | 0.018 | 0.55 | Rendering | 1.625 | 11.373 | 7.528 | <0.001 | 0.518 |
| Environment × Rendering | 2.19 | 15.33 | 38.54 | <0.001 | 0.85 | Environment × Rendering | 8 | 56 | 10.511 | <0.001 | 0.6 |
| **Envelopment** | | | | | | **Envelopment** | | | | | |
| Pmode | 1 | 7 | 13.17 | 0.008 | 0.65 | | | | | | |
| Pmode × Rendering | 1.02 | 7.16 | 8.56 | 0.021 | 0.55 | | | | | | |
| Environment × Rendering | 1.81 | 12.64 | 4.2 | 0.043 | 0.38 | | | | | | |
| **Metallic Tone Color** | | | | | | **Metallic Tone Color** | | | | | |
| Environment | 2 | 14 | 4.74 | 0.027 | 0.40 | Environment | 2 | 14 | 11.27 | =0.001 | 0.62 |
| Rendering | 1.33 | 9.28 | 5.27 | 0.039 | 0.43 | Environment × Rendering | 2.27 | 15.9 | 25.56 | <0.001 | 0.79 |
| Environment × Rendering | 1.72 | 12.01 | 47.08 | <0.001 | 0.87 | | | | | | |
| **Reverberation** | | | | | | **Reverberation** | | | | | |
| Environment | 2 | 14 | 29.45 | <0.001 | 0.81 | Environment | 2 | 14 | 17.923 | <0.001 | 0.72 |
| Rendering | 1.32 | 9.23 | 5.35 | 0.038 | 0.43 | Rendering | 1.52 | 10.63 | 9.81 | 0.006 | 0.58 |
| Pmode × Environment | 2 | 14 | 6.6 | 0.01 | 0.49 | | | | | | |
| Pmode × Rendering | 1.16 | 8.1 | 6.93 | 0.027 | 0.5 | | | | | | |
| Environment × Rendering | 1.81 | 12.66 | 25.84 | <0.001 | 0.79 | Environment × Rendering | 2.13 | 14.9 | 8.75 | 0.003 | 0.56 |
| **Tone Color** | | | | | | **Tone Color** | | | | | |
| Environment | 2 | 14 | 4.73 | 0.027 | 0.40 | | | | | | |
| **Source Width** | | | | | | **Source Width** | | | | | |
| Environment | 2 | 14 | 13.9 | <0.001 | 0.67 | | | | | | |
| Pmode × Environment | 2 | 14 | 4.62 | 0.012 | 0.40 | | | | | | |
| Environment × Rendering | 1.94 | 13.61 | 29.13 | <0.001 | 0.81 | | | | | | |
| **Externalization** | | | | | | **Externalization** | | | | | |
| Pmode | 1 | 7 | 9.39 | 0.018 | 0.57 | Environment | 1.11 | 7.8 | 6.012 | 0.038 | 0.46 |
| Rendering | 2 | 14 | 8.04 | 0.005 | 0.54 | Rendering | 3 | 21 | 24.943 | <0.001 | 0.78 |
| Environment × Rendering | 1.55 | 10.88 | 8.11 | 0.010 | 0.54 | Environment × Rendering | 6 | 42 | 11.601 | <0.001 | 0.62 |
| Stimulus × Environment | 2 | 14 | 3.95 | 0.044 | 0.36 | | | | | | |
| Pmode × Rendering | 2 | 14 | 5.3 | 0.019 | 0.43 | | | | | | |
| **Overall Difference** | | | | | | | | | | | |
| Stimulus | 2 | 14 | 60.63 | <0.001 | 0.90 | | | | | | |
| Environment | 2 | 14 | 237.90 | <0.001 | 0.97 | | | | | | |
| Rendering | 1.61 | 11.23 | 470.88 | <0.01 | 0.99 | | | | | | |
| Pmode × Stimulus | 2 | 14 | 7.74 | 0.005 | 0.53 | | | | | | |
| Pmode × Environment | 2 | 14 | 6.51 | 0.010 | 0.48 | | | | | | |
| Stimulus × Environment | 4 | 28 | 9.99 | <0.001 | 0.59 | | | | | | |
| Stimulus × Rendering | 6 | 42 | 118.28 | <0.001 | 0.94 | | | | | | |
| Environment × Rendering | 6 | 42 | 110.65 | <0.001 | 0.94 | | | | | | |
| **Speech Intelligibility** | | | | | | | | | | | |
| Environment | 1.20 | 8.44 | 54.081 | <0.001 | 0.89 | | | | | | |
| Rendering | 3 | 21 | 146.483 | <0.001 | 0.95 | | | | | | |
| Environment × Rendering | 2.36 | 16.54 | 12.712 | <0.001 | 0.65 | | | | | | |
| **Plausibility** | | | | | | | | | | | |
| Rendering | 1.03 | 7.24 | 34.08 | <0.001 | 0.83 | | | | | | |
| Stimulus × Rendering | 1.21 | 8.53 | 28.31 | <0.001 | 0.80 | | | | | | |
| Environment × Rendering | 4 | 28 | 3.75 | 0.014 | 0.35 | | | | | | |